\documentstyle[titlepage]{article}
\newcommand{\R}{{\rm I\kern-2pt R}}

\newtheorem{theorem}{Theorem}[section]
\newtheorem{remark}{Remark}[section]

\begin{document}
\begin{center}
{\bf Control of a  Coupled Two Spin System Without Hard Pulses\\}
Viswanath Ramakrishna\footnote{Corresponding Author} \\
Department of Mathematical Sciences
and Center for Signals, Systems and Communications\\
University of Texas at Dallas\\
Richardson, TX 75083 USA\\
email: vish@utdallas.edu\\
Supported in Part by NSF -DMS 0072415\\ 
Raimund J. Ober\\
Department of Electrical Engineering
and Center for Signals, Systems and Communications\\
University of Texas at Dallas\\
Richardson, TX 75083, USA\\
email: ober@utdallas.edu\\
Supported in Part by NSF-DMS 9803186\\ 
Kathryn L. Flores\\
Department of Mathematical Sciences
and Center for Signals, Systems and Communications\\
University of Texas at Dallas\\
P. O. Box 830688\\
Richardson, TX 75083 USA\\
email: kflores@utdallas.edu\\
Herschel Rabitz\\
Department of Chemistry\\
Princeton University\\
Princeton, NJ 08544\\
email: hrabitz@chemvax.princeton.edu\\
Supported in Part by DOD and NSF  

\end{center}

\begin{abstract}
Constructive techniques for controlling a coupled, heteronuclear spin system,
via bounded amplitude sinusoidal pulses are presented. The technique
prepares exactly any desired unitary generator in a rotating frame, through
constant controls.
Passage to the original coordinates provides a procedure
to prepare arbitrary unitary generators, via bounded amplitude
piecewise sinusoidal
pulses whose frequency is one of the two Larmor frequencies and whose
phase takes one of two values. The techniques are based on a certain
Cartan decomposition of $SU(4)$ available in the literature. A method 
for determining the parameters entering this Cartan decomposition,
in terms of the entries of the target unitary generator, is also provided.
\end{abstract}

\begin{section}{Introduction}
The goal of this paper is to provide, constructively and exactly
(i.e., without any approximations) a decomposition:
\begin{equation}
\label{Maindecomp}
e^{-iLI_{ij}} = \Pi_{k=1}^{Q}e^{(-ia_{k}I_{1z}I_{2z} -ib_{k}I_{ij})}, i=1,2,
j =x,y  
\end{equation}
satisfying i) {\bf O1} $a_{k} > 0$; and ii) {\bf O2} $\mid b_{k} \mid 
\leq C$, for some prescribed bound $C$. The matrices, $I_{ij},
i=1,2, j = x,y,z$, defined in Section 2, are the standard tensor
products of $I_{2}$ and the Pauli matrices. As will become clear,
later the same
decomposition can be refined to
simultaneously satisfy $\mid \frac{b_{k}}{a_{k}}\mid \leq D$ for another
bound $D$. This problem arises in the control of a  coupled, heteronuclear,
two spin system which is being controlled by addressing each spin individually.
The adjective ``heteronuclear" is , throughout this paper,
meant only to signify that
the Larmor frequencies of the two spins are different.
The condition {\bf O1} just means that the time for which a pulse has to
be applied must necessarily positive, while {\bf O2} (or more precisely,
the condition $\mid \frac{b_{k}}{a_{k}}\mid \leq D$) represents the fact
that {\it hard pulses are not being used.}

Coupled spin systems are useful from several points of view. They are
certainly ubiquitous in NMR studies, \cite{palmer,slichter}. They also
provide examples of coupled qubits in quantum computation and 
information processing, \cite{divince1,divince2,havel,ike}. One method to
control spin systems is to use hard pulses, i.e., high amplitude pulses
applied for very short times.  
The usage of hard pulses,
though useful from the perspective of minimizing the time consumed to
prepare a desired unitary generator, $S$,
has its own problems. Perhaps \underline {most} importantly,
the usage of hard pulses allows
one to only approximately prepare $S$,
since the available resources in
any situation is limited. For certain problems of NMR spectroscopy this may
not be such a big restriction since the translation of a certain NMR objective
to the problem of unitary generator preparation may allow for a broad choice
of unitary generators (though, to the best
of our knowledge, there has been no systematic assessment of the inaccuracies
introduced by the usage of hard pulses for specific NMR objectives).
However, for purported quantum computation applications it is indeed a
restriction, since the 
desired $S$ has to be obtained with great accuracy.
Secondly, the usage of hard pulses may violate
the basic feature assumed often in the methodology - namely the ability to
address single spins selectively. This already necessitates restriction
to heteronuclear molecules. However, even for heteronuclear molecules
the usage of infinite amplitude is problematic. Indeed, neglecting modes
which are coupled to a system being studied is essentially a perturbation
theory argument. Thus, not only has the frequency of the pulse got to be
in resonance with the subsystem being studied, but the pulse area has
also got to be bounded. If high amplitudes are being used then the 
pulse has to be applied for an extremely short time. It is unrealisitic
to assume that a pulse can be applied for an infinitesimal time in
the laboratory.
The situation
is analogous to the usage of the rotating wave approximation in molecular
control studies, \cite{herschnew,pranew,atom}. Not only must there be
no resonances amongst the coupled levels, but the amplitude of the pulse
must be much smaller than the frequency separation between the pair
of levels being addressed.
Similar considerations occur in NMR spectroscopy, \cite{pound}.
These problems are compounded
further in quantum computation applications, since it is desirable to
perform local operations on individual (or small collections of) qubits,
without too much crosstalk with other qubits during these operations.
Usage of hard pulses, will eventually cause other qubits to be coupled.
Put differently, one of the goals of hard pulse technology, namely the
avoidance of decoherence effects, can in fact be defeated by the very usage
of hard pulses.
Thus, while it is desirable to finish all control action
before relaxation processes become dominant, it is even more 
important to use fields which do not cause other couplings or processes
neglected in
a model to become significant. Quite often this means essentially that
the field be bounded in amplitude. 
{\it In this paper, arbitrary bounds on the amplitude will be allowed.
Thus, while $\omega_{1}\neq\omega_{2}$ (where the $\omega_{i}, i=1,2$ are
the individual frequencies) is needed, the difference need not
be as large, as would be required by the usage of hard pulses.}
Thus, there is lots to be said for controlling spin systems 
{\it without} hard pulses. In this paper, explicit techniques which avoid
hard pulses are provided for a coupled spin system.

It is worthwhile to place in proper mathematical
context the question being studied.   
Coupled two spin systems can be
viewed as examples of left-invariant systems, with drift,
evolving on the compact Lie group, $SU(4)$, \cite{sussmann}. It is known
{\bf non-constructively} that, under a certain Lie algebraic condition,
such systems can be controlled with piecewise
constant pulses whose amplitude can be arbitrarily bounded \cite{sussmann}.
Providing constructive proofs for such results, on the other hand,
is an entirely different matter. There are two sources of complication.
First, the ambient
Lie group, $SU(4)$, is high-dimensional (precisely, fifteen dimensional).
But more importantly, the presence of a drift term (i.e., the free Hamiltonian)
significantly complicates constructive control. Indeed, from a control
theoretic perspective hard pulse arguments are ways of avoiding the
effects of drift (though at the expense of introducing far
more deleterious effects). There is an extensive literature on constructive
control for driftless systems - see, for instance, the survey, \cite{mcclam}.
The same reference contains a survey of special classes of classical
mechanical systems with drift, which can be controlled constructively.
In \cite{gnvn99an7103}, building on earlier work \cite{pranew}, a detailed
study of constructive control of systems with drift on $SU(2)$ was provided.
The special structure of $SU(2)$ played a crucial role in this effort.
In this paper,
we will show that the special structure of $SU(2)$ allows, once again,
to constructively control the spin system being studied here.
This structure of $SU(2)$ enters in two manners.
First, the groups $SU(4)$ and $SO(4)$ are intimately connected
to one another due to one well-known Cartan decomposition of $SU(4)$.  
This Cartan decomposition factors every $S\in SU(4)$ as products of matrices
in $SO(4)$ and matrices which are exponentials of a Cartan subalgebra
of the complement of $so(4)$, the Lie algebra of $SO(4)$,
in $su(4)$, the Lie algebra of $SU(4)$ \cite{gilmore,hermann,weaver}.
Now $SO(4)$ is essentially 
the same as $SU(2)\otimes SU(2)$. Not only does this fact clearly 
establish this Cartan decomposition (this is well known),
but it also enables techniques inspired
by our earlier paper, \cite{gnvn99an7103}, for systems on $SU(2)$ to
achieve the main goals of this paper.
The second role of $SU(2)$ is to facilitate  
the calculation of the real parameters in this Cartan decomposition. 
In the appendix, a method to determine these parameters
as explicitly as possible (explicit modulo the solution of a
pair of transcendental equations) is provided, and once again
the structure of
$SU(2)$ is the main ingredient.

In \cite{thieves} the authors use this particular Cartan decomposition
of $SU(4)$ to address the problem of generating any unitary generator
$S\in SU(4)$ for a coupled two particle spin system
via selective one-spin \underline{hard} pulses. 
The essential difference
between the approach taken here and that in \cite{thieves}
stems from a {\it Lie theoretic nuance} and
may be summarized as follows. This Cartan decomposition and plus
a few calculations leads to a factorization
of any $S\in SU(4)$, \cite{thieves}:  
\begin{equation}
\label{shorteuler} 
S = \Pi_{k=1}^{Q} {\mbox exp} (-it_{k}A_{k})
\end{equation}
where each $A_{k}$ is one of $I_{1z}I_{2z}, I_{1x}, I_{2x}, I_{1y}, I_{2y}$.
Greater detail about the decomposition, (\ref{shorteuler}), is provided
in Equation (\ref{ourdecomp}) in the next section.
Under the assumption that it is possible to address single spins selectively,
(and a subsequent passage to a rotating frame) $I_{1z}I_{2z}$ represents
the internal Hamiltonian, whereas the remaining
matrices represent the control coupling. In \cite{thieves} factors of the
type ${\mbox exp} (-it_{k}I_{iz}I_{2z})$ are generated by free evolution,
whereas factors which are exponentials of the control couplings are generated
by hard pulses, so that the drift term, i.e., $I_{1z}I_{2z}$, makes a
negligible contribution to such a factor. This, naturally, makes
the generation of
the desired $S$ (even in the rotating frame) at best approximate. 
In our approach, we
will also generate the factors ${\mbox exp} (-it_{k}I_{iz}I_{2z})$ through
free evolution. {\it However, instead of viewing the remaining factors as the
exponentials of the control couplings, we view them as the exponentials
of the iterated commutator of
the internal Hamiltonian and the control couplings.} In this iterated
commutator, the internal Hamiltonian occurs twice whereas the control
couplings occur only once.
Thus it may be surmised, based on our prior experience \cite{gnvn99an7103},
that such exponentials
{\it can be generated via three factors, 
two of which are free evolution terms and
one is a control (though not hard) pulse.} 
While, this conjecture has not been established for general nonlinear
control systems, 
it turns out to be valid for the two spin
system studied in this paper. Put differently, in generating the 
exponentials of the control couplings we make systematic use of the
drift term, as opposed to treating it as a nuisance which can be overcome
via hard pulses. As a further step towards making the
whole procedure constructive methods which yield the $t_{k}$
in Equation (\ref{shorteuler}) in
terms of the entries of the target unitary generator, $S$, will be displayed.
The fact that the Cartan decomposition is
the analogue of the Euler decomposition for the higher dimensional 
unitary groups is well known, \cite{gilmore,hermann,weaver}.
However, a stumbling
block in its usage is the fact that the analogues of the Euler angles, i.e.,
the real parameters $t_{k}$, 
have never been explicitly written down in terms of the entries of the
target matrix $S$.  

Ultimately, the choice of hard versus soft pulses depends on both the
system and the target in question. The contribution of this paper is to
show that many objectives can be met via soft pulses. Furthermore, this is
demonstrated by making use of the iterated commutators of the internal
Hamiltonian and the control Hamiltonians, in contrast to hard pulse
approaches which make no use of this commutator. This is another reason
why the results below may be of interest. Indeed, all arguments for
the controllability of
finite dimensional quantum systems \cite{sussmann,turnici,sonia,pra,sally1,jmc} 
hinge on the
commutators of the internal Hamiltonian with the control Hamiltonians.
Further, these arguments also show that it is possible (non-constructively)
to prepare any target via bounded amplitude fields. Thus, it is at least
didactically pleasing to demonstrate controllability constructively
via methods which
explicitly use such iterated commutators.  
In summary, two important problems are solved constructively in
this paper: a) the control of a coupled spin system via sinusoidal,
bounded amplitude pulses assuming the factors $t_{k}$ of
Equation (\ref{shorteuler}) are given, and b) finding these parameters,
$t_{k}$, algorithmically (so that there is something useful even for
afficionados of hard pulses).
For both these problems a significant role
is played by the structure of $SU(2)$.

The balance of this paper is organized as follows. In the next section
some notation and basic facts about the Lie group $SU(2)$ will be collected.
In particular, an explicit formula for an
Euler angle factorization, with factors exponentials of $\sigma_{x}$
and $\sigma_{y}$, is provided. The third section carefully derives the
rotating frame for the basic model, and shows how to determine the
frequencies and phases of the piecewise sinusoidal pulses which 
will be designed to generate any $S\in SU(4)$.
The fourth section describes how to determine the amplitudes of 
these pulses and in the process explains how the two goals,
{\bf O1} and {\bf O2}
mentioned at the very begining of this section, help in finding the amplitude
and the duration of the pulses.
Some conclusions are offered in the next section. An appendix describes 
how to go about calculating the Euler angle analogues for $SU(4)$.   

\end{section}
\begin{section} {Review of Basic $SU(2)$ Facts}
The Pauli matrices will be denoted as:
 $\sigma_{x} =
 \left (\begin{array}{cc}
 0 & 1\\
 1 & 0
 \end{array}
 \right)$, $\sigma_{y} =
 \left (\begin{array}{cc}
 0 &  -i\\
 i & 0
 \end{array}
 \right )$ and $\sigma_{z}
 = \left (\begin{array}{cc}
 1 & 0\\
 0 & -1
 \end{array}
 \right)$
In terms of this definition of the Pauli matrices, a few important
$4\times 4$ Hermitian matrices can also be defined:
\begin{equation}
I_{1k} = \sigma_{k}\otimes I_{2}; I_{2k} = I_{2}\otimes \sigma_{k},
\ k = x,y,z; 
\end{equation}
Note that both in the definition of the Pauli matrices and in the above
equation, the customary factor of $\frac{1}{2}$ has been omitted. This is
for notational convenience and does not affect any of the results below.

Next, a very useful representation of the $SU(2)$ matrices, 
the Cayley-Klein parametrization, follows. 

\begin{equation}
\label{specialuseful}
S = S (\alpha , \zeta , \mu) =  \left ( \begin{array}{cc}
e^{i\zeta }\cos \alpha & e^{i\mu}\sin \alpha\\
e^{i(\pi - \mu )}\sin \alpha & e^{-i\zeta} \cos \alpha
\end{array}
\right )
\end{equation}
$\alpha , \zeta , \mu$ are the Cayley-Klein parameters of $S$.
Since this parametrization is nothing but the entries
of $S$ written in polar form, it is clear that
$\zeta$ and $\mu$ may be taken to be in $[0, 2\pi)$ and
$\alpha$ to be in $[0, \frac{\pi}{2}]$.

Crucial for the purposes of this paper is an Euler parametrization,
in terms of $\sigma_{x}$ and $\sigma_{y}$, of any $S(\alpha , \zeta , \mu )$:

\begin{equation}
S(\alpha , \zeta , \mu ) = e^{iD\sigma_{x}}e^{iE\sigma_{y}}e^{iF\sigma_{x}}
\end{equation}
Though, in principle, the Euler angles $D, E$ and $F$ may be obtained
from more common Euler angle parametrizations \cite{harter},
an explicit formula for
them in terms of the Cayley-Klein coordinates goes a long way towards
making the techniques of this paper genuinely constructive. Therefore, the
following relations are very useful, \cite{gnvn99kathryn}:

\begin{eqnarray}
\label{eulerangles}
\cos (E) & = & \sqrt{\cos^{2}\zeta\cos^{2}\alpha + \sin^{2}\mu\sin^{2}\alpha }\\
\sin (D - F) & = & \frac{\sin\zeta\cos\alpha}
{\sqrt{\sin^{2}\zeta\cos^{2}\alpha + \cos^{2}\mu\sin^{2}\alpha }}\\
\sin (D + F) & = & \frac{\sin\mu\cos\alpha}   
{\sqrt{\sin^{2}\mu\sin^{2}\alpha + \cos^{2}\zeta\cos^{2}\alpha }}
\end{eqnarray} 
The parameters, $D, E$ and $F$ can be assumed to be in $[0, 2\pi)$
if needed.

The next item on the list is the following decomposition of any matrix,
$S\in SU(4)$ into factors which are exponentials of $I_{ij}, i=1,2, j =x,y$
and the matrix $I_{1z}I_{2z}$, \cite{thieves}.

\begin{equation}
\label{highwayrobbery}
S = K_{1}\otimes K_{2}
e^{-i\frac{\pi}{4}I_{1y}}e^{-i\frac{\pi}{4}I_{2y}}
e^{-i\theta_{1}I_{1z}I_{2z}}   
e^{-i\frac{7\pi}{4}I_{1y}}e^{-i\frac{7\pi}{4}I_{2y}}
e^{-i\frac{7\pi}{4}I_{1x}}e^{-i\frac{7\pi}{4}I_{2x}}
e^{-i\theta_{2}I_{1z}I_{2z}}
e^{-i\frac{\pi}{4}I_{1x}}e^{-i\frac{\pi}{4}I_{2x}}
e^{-i\theta_{3}I_{1z}I_{2z}}
K_{3}\otimes K_{4}
\end{equation}
In (\ref{highwayrobbery}) the matrices $K_{i}, i = 1, \ldots, 4$ are some
matrices in $SU(2)$. This decomposition follows from the well known fact
that the Lie algebra $su(2)\otimes su(2)$ and its orthogonal complement
in the Lie algebra $su(4)$ provide a Cartan decomposition of the
Lie group $SU(4)$, \cite{hermann,gilmore,weaver}.  
In \cite{thieves} this fact and some calculations are used to obtain
the decomposition (\ref{highwayrobbery}). {\it Note, however that
they do not provide any formulae for the $K_{i}, i =1, \ldots, 4$ 
and the $\theta_{k}, k=1, \ldots , 3$ in terms of the target matrix $S$.}
In the appendix, we will ameliorate
this problem. For the moment, however, expanding each of the $K_{i}, i =1,
\ldots , 4$ into its $\sigma_{x}, \sigma_{y}$
Euler angles via Equation (\ref{eulerangles}) and using some Kronecker calculus,
leads to the following equation, which is the one we will work with:
\begin{eqnarray}
\label{ourdecomp}
S  & = & e^{iD_{1}I_{1x}}e^{iE_{1}I_{1y}}e^{iF_{1}I_{1x}}
e^{iD_{2}I_{2x}}e^{iE_{2}I_{2y}}e^{iF_{2}I_{2x}}
e^{-i\frac{\pi}{4}I_{1y}}e^{-i\frac{\pi}{4}I_{2y}}
e^{-i\theta_{1}I_{1z}I_{2z}}\\ \nonumber  
& \ & e^{-i\frac{7\pi}{4}I_{1y}}e^{-i\frac{7\pi}{4}I_{2y}}
e^{-i\frac{7\pi}{4}I_{1x}}e^{-i\frac{7\pi}{4}I_{2x}}
e^{-i\theta_{2}I_{1z}I_{2z}}\\ \nonumber 
& \ & e^{-i\frac{\pi}{4}I_{1x}}e^{-i\frac{\pi}{4}I_{2x}}
e^{-i\theta_{3}I_{1z}I_{2z}}
e^{iD_{3}I_{1x}}e^{iE_{3}I_{1y}}e^{iF_{3}I_{1x}}   
e^{iD_{2}I_{2x}}e^{iE_{2}I_{2y}}e^{iF_{2}I_{2x}}
\end{eqnarray}

The $(D_{i}, E_{i}, F_{i})$
are the Euler angles of the $K_{i}\in SU(2), i=1, \ldots ,  4$.  
\end{section}

\begin{section} {Determination of the Frequencies and Phases}
Consider a pair of coupled spins in the weak coupling limit.
The system is assumed to be heteronuclear, so that it is possible
to address each spin individually as long the as the frequency of the
corresponding field is resonant with the Larmor frequency of
the spin, $\omega_{i}$, in question and the field
can be bounded in amplitude. 
Thus, the model is

\[
\dot{V} = -\frac{i}{2}(\hat{A}V + u_{1}(t)B_{1}V + u_{2}(t)B_{2}V), V\in SU(4) 
\] 
where the internal Hamiltonian,
$\hat{A}$ is $\omega_{1}\sigma_{z}\otimes I_{2} +
 \omega_{2}I_{2}\otimes\sigma_{z} +
J \sigma_{z}\otimes \sigma_{z}$, and the interaction Hamiltonians are
$B_{1} = b_{1}\sigma_{x}\otimes I_{2}$ and $B_{2} =
b_{1}\sigma_{y}\otimes I_{2}$. The $b_{i}, i=1,2$ ($b_{2}$ appears below) 
are constants related to the gyromagnetic ratios, and the $\omega_{i}, 
i = 1,2$ are the Larmor frequencies.
Finally, $u_{1}(t)$ and $u_{2}(t)$ are
sinusoidal fields to be designed:
\begin{equation}
\label{ober}
u_{1}(t) = c\cos (\omega t + \phi ),
u_{2} = c\sin (\omega t + \phi )
\end{equation} 
The frequency, $\omega$, will be taken
to be $\omega_{1}$. So the design procedure amounts to specifying
the amplitude, $c$, the phase, $\phi$ and the duration of the pulses.
These will be chosen in a piecewise constant manner.
The fourth section is essentially devoted to finding the amplitudes and
the durations.
How the phases ought to be chosen will become clear later in this section.

The above equation was derived assuming that the first spin was 
being addressed. If the second spin is being addressed, then $B_{1},
B_{2}$ would be replaced by $b_{2}I_{2}\otimes \sigma_{x}$ 
and $b_{2}I_{2}\otimes\sigma_{y}$ respectively, while the frequency,
$\omega$, of the field would be replaced by $\omega_{2}$.

Let us now derive the rotating frame in which the problem of
preparation of a target will be translated into finding picewise
constant controls for an associated system in the rotating frame
(see \cite{sally2} for related considerations).
Set 
\[
U(t) = e^{tF}V(t)
\] with 
\[
F = \frac{i}{2}(\omega_{2}I_{2}\otimes\sigma_{z}
+ \omega_{1}\sigma_{z}\otimes I_{2})  
\]

Then a few calculations reveal that 
\begin{equation}
\label{firststop}
\dot{U} = -\frac{i}{2}(JI_{1z}I_{2z})U - \frac{i}{2}
(cb_{1}\Delta\otimes I_{2})U 
\end{equation}

where
\[
\Delta = \left ( \begin{array}{cc}
0 & e^{-i\phi } \\
e^{i\phi } & 0
\end{array} \right )
\]

Note that the matrix $\Delta$ is {\it independent of time and is parametrized
by the phase, $\phi$, of the field which can be choosen.}
Thus, if we choose $\phi = 0$, $\Delta$ is $\sigma_{x}$ and if
we set $\phi = \frac{\pi}{2}$, $\Delta$ is $\sigma_{y}$.  

A similar calculation reveals that the {\bf same rotating frame} can be used
to address the other spin (with the frequency of the field,
$\omega = \gamma_{2}$) to
obtain the following equation
\begin{equation}
\dot{U} = -\frac{i}{2}(JI_{1z}I_{2z})U 
-\frac{i}{2}(cb_{2}I_{2}\otimes\Delta ) U 
\end{equation}

Once again by choosing the phase, $\phi$, one can ensure that
$\Delta$ is either $\sigma_{y}$ or $\sigma_{x}$.

{\it The upshot} of the foregoing is that by choosing the frequency of
the field to be resonant with one of the spins and by choosing the
phase in an appropriate manner, passage to a unique rotating frame
leads to the following system, which is controlled by constant inputs:

\begin{equation}
\label{inrframe}
\dot{U} = -\frac{i}{2}AU  -\frac{i}{2}dBU, U\in SU(4)
\end{equation}
with $A = J\sigma_{z}\otimes \sigma_{z}$ and $B$ one of the matrices
$I_{1x}, I_{2x}, I_{1y}, I_{2y}$ where $I_{1j} = \sigma_{j}\otimes I_{2},
j = x, y$ and $I_{2j} = I_{2}\otimes \sigma_{j}, j =x, y$.   
The constant, $d$, is related to the amplitude of the field and other
constants of the system. This is useful because the Cartan decomposition, 
i.e., Equation (\ref{ourdecomp}) of Section 2,
consists precisely of the exponentials of one of $A, I_{ij}, i= 1, 2, 
j= x, y$. Thus, if the exponential of a certain $I_{ij}$ is required we
choose the frequency and phase of the field so that $B$ of
Equation (\ref{Unitary}) becomes $I_{ij}$, and then follow the procedure
in the next section to determine, $d$ (hence, the amplitude of the field).

It is interesting to observe that while in the rotating frame, we are
exciting one spin with one of the $x$ or $y$ magnetic field components,
in the original frame we are exciting any one spin by using {\it both}
the  $x$ and $y$ magnetic field components.

\end{section}

\begin{section} {Determining the Amplitudes} 
The basic model derived in the previous section leads
to the following system in the rotating frame, for which we will
design piecewise constant controls:

\begin{equation}
\label{Unitary}   
\dot{U} = -iAU - iu(t)BU, \ U(0) = I_{4}
\end{equation}
where $A = I_{1z}I_{2z} = \sigma_{z}\otimes\sigma_{z}$,
and $B$ is one of the $I_{ij}, i =1,2, j = x,y$ and $u(t)$ the control
(to be determined) is piecewise constant.

{\it Note, for the sake of easy bookeeping,
the  constants $J$, $b_{i}$  and
$\frac{1}{2}$ have been dropped in
Equation (\ref{Unitary}). However, once results for the above
model are available it is a routine matter to derive results for
the actual model, with these parameters present.}

Going back to the equation (\ref{ourdecomp}) it is clear that 
factors which are the exponentials of $-iI_{1z}I_{2z}$ (i.e., of $-iA$) 
can be generated by free evolution
i.e., by setting $u(t) = 0$ for an amount of time given by the
corresponding $\theta_{k}$.
Indeed, $e^{-i\theta_{k}A}$ is a diagonal matrix, 
${\mbox diag} (e^{-i\theta_{k}},
e^{i\theta_{k}}, e^{i\theta_{k}}, e^{-i\theta_{k}})$.
If $\theta_{k} \geq 0$, then
$e^{-i\theta_{k}A}$ can be prepared
by free evolution for $\theta_{k}$ units of time.
If $\theta_{k} < 0$ then, $e^{-i\theta_{k}A}$
can be prepared by free evolution
for $2\pi + \theta $ units of time. {\bf Only for free evolution terms
are periodicity arguments resorted to.}  For other factors, instead of
resorting to periodicity arguments, substantial use of the structure
of the drift $-iA$
will be made.

So all that remains to be addressed is preparing factors which are
exponentials of the $-iI_{ij}$ via controlled pulses. In other words,
the main goal at this stage
is to decompose the exponential, $e^{-iLI_{ij}}, i=1,2, j = x,y$,
for any $L\in R$, as:   

\begin{equation}
e^{-iLI_{ij}} = \Pi_{k=1}^{Q} e^{(-ia_{k}A  -ib_{k}I_{ij})}, i=1,2,
j=x,y, \ A = I_{1z}I_{2z}   
\end{equation}
satsifying i) {\bf O1} $a_{k} > 0, k =1, \ldots , Q$ and 
ii) {\bf O2} $\mid b_{k}\mid \leq C, k =1, \ldots , Q$.
Constructive methods for obtaining the desired $a_{k}, b_{k}$ in
terms of $S$ will be provided. At this stage the formulae for the $a_{k}$
and $b_{k}$ will be provided in terms of the parameters, $(D_{i}, E_{i},
F_{i}), i=1, \ldots , 4$ and $\theta_{k}, k =1, \ldots, 3$ of 
Equation (\ref{ourdecomp}) 
The appendix indicates a procedure which
obtains these $SU(4)$ ``Euler angles" directly from the entries
of $S$. Note that $a_{k}$ is the duration of the  $k$th pulse 
and $\frac{b_{k}}{a_{k}}$ is its amplitude.
The basic strategy is to first achieve
{\bf O1} and then constructively modify the resulting decomposition to
meet {\bf O2} also. The proof below shows how achieving {\bf O2} 
{\it leads also to arbitrarily bounding the amplitude of the pulse.}

\begin{theorem}
\label{keytheorem}
{\rm The matrix ${\mbox exp}(-iL I_{1x} ), L\in R$ can be factored
explicitly as $\Pi_{k=1}^{3} {\mbox exp} (-ia_{k}A  - ib_{k}I_{1x})$
with $A = I_{1z}I_{2z}$ and with $a_{k} > 0, k=1, \ldots , 3$
and $b_{1} = 0 = b_{3}$, if $\cos L \neq 0$.
If $\cos L = 0$, then four factors are needed, with $b_{1} = 0 = b_{4}$.
Further, this factorization can be refined constructively,
by increasing the number of factors, to ensure that $\mid b_{k}\mid
\leq C, k =1, \ldots , Q$. This decomposition can be further refined to
meet the condition, $\mid \frac{b_{k}}{a_{k}}\mid\leq D$, for any
prescribed $D > 0$.   Similar statements hold for the exponentials
of $I_{2x}, I_{1y}$ and $I_{2y}$ with the matrix 
for appropriately different sets of $a_{k}$'s and $b_{k}$.} 
\end{theorem}

{\bf Proof:} The proof will be given only for the exponential 
of ${\mbox exp}(-iL I_{1x} )$, since the proof for the others are
very similar.

The special structure of $I_{1z}I_{2z}$ and $I_{1x}$ results in the
following matrix for ${\mbox exp} (-ia_{k} A  -ib_{k} I_{1x})$ 

\begin{equation}
\left ( \begin{array}{cccc}
\cos \lambda_{k} - \frac{i}{\lambda_{k}}a_{k}\sin \lambda_{k} &
0 &\frac{-ib_{k}}{\lambda_{k}} \sin \lambda_{k} & 0 \\
0 & 
\cos \lambda_{k}  + \frac{i}{\lambda_{k}}a_{k} \sin \lambda_{k} &
0 &\frac{-ib_{k}}{\lambda_{k}} \sin \lambda_{k} \\
\frac{-ib_{k}}{\lambda_{k}}\sin \lambda_{k} & 0
& \cos \lambda_{k} + \frac{i}{\lambda_{k}}a_{k} \sin \lambda_{k} &
0 \\
0 &\frac{-ib_{k}}{\lambda_{k}} \sin \lambda_{k} &
0 &
\cos \lambda_{k} - \frac{i}{\lambda_{k}}a_{k} \sin \lambda_{k}  
	\end{array} \right )
\end{equation} 

where $\lambda_{k} = \sqrt{(a_{k}^{2} + b_{k}^{2}}$. In particular,
setting $a_{k} = 0$ and $b_{k} = L$ yields the exponential
of $-iL I_{1x}$. Similarly, an explicit calculation yields the
following formula:

\begin{equation}
{\mbox exp} (-iL \sigma_{y}\otimes \sigma_{z})
=\left ( \begin{array}{cccc}
\cos L &  0 & \sin L  & 0 \\
0 & \cos L & 0 & -\sin L   \\
-\sin L & 0 & \cos L & 0\\   
0 & \sin L & 0 & \cos L 
\end{array} \right )    
\end{equation}

Using the last two formulae, the following useful identity is obtained:
\begin{equation}
{\mbox exp} (-iL I_{1x}) = {\mbox exp} (-i\frac{7\pi}{4}I_{1z}I_{2z})
{\mbox exp} (-iL \sigma_{y}\otimes\sigma_{z})
{\mbox exp} (-i\frac{\pi}{4}I_{1z}I_{2z}) 
\end{equation}
Thus, to prepare the matrix ${\mbox exp} (-iL I_{1x})$, free evolution
for $\frac{7\pi}{4}$ units of time for the first factor above is used and 
free evolution for $\frac{\pi}{4}$ units of time  for the third
factor is used. 
Therefore, it remains to produce the middle factor, 
${\mbox exp} (-iL \sigma_{y}\otimes\sigma_{z})$ via controlled pulses.
Since, the 
matrix $-iL \sigma_{y}\otimes\sigma_{z}$ is, upto a constant,
the commutator of
$A$ and $B_{1}$, it seems plausible that its exponential can be represented as
a product, $\Pi_{k=1}^{2} {\mbox exp} [-i(a_{k}I_{1z}I_{2z} + b_{k}I_{1x})]$.

To demonstrate this, three cases need to be considered:
i) $\cos L >  0$, ii) $\cos L < 0$ and
iii) $\cos L = 0$. 
 
\noindent {\bf The Case $\cos L > 0$:} Evaluating the product 
$\Pi_{k=1}^{2} {\mbox exp} [-i(a_{k}I_{1z}I_{2z} + b_{k}I_{1x})]$
and choosing $\lambda_{1} = \sqrt{a_{1}^{2} + b_{1}^{2}}
= \frac{3\pi}{2}$ and  $\lambda_{2} = \sqrt{a_{2}^{2} + b_{2}^{2}} 
= \frac{\pi}{2}$ and equating the result to 
the matrix ${\mbox exp} (-iL \sigma_{y}\otimes\sigma_{z})$ 
leads to the following equations:
\begin{eqnarray*}
\frac{a_{1}a_{2} + b_{1}b_{2}}{\lambda_{1}\lambda_{2}}
& = & \cos  L \\
\frac{a_{1}b_{2} -  a_{2}b_{1}}{\lambda_{1}\lambda_{2}} &
= & -\sin  L     
\end{eqnarray*}
These two equations can be solved as follows. Choose $b_{1} = 0,
a_{1} = \frac{3\pi}{2}, 
b_{2} = -\frac{\pi}{2}\sin   L,
a_{2} = \frac{\pi}{2}\cos  L$.  
Thus, {\bf 01} has been
met for the case that $\cos \ L > 0$. 
 
\noindent {\bf The Case $\cos  L < 0$:} Now choose $\lambda_{1}
= \frac{\pi}{2}$ and $\lambda_{2} = \frac{\pi}{2}$.  
The resulting set of equations has the following solution: 
$b_{1} = 
 - \frac{\pi}{2}\sin \ L, a_{1} = - \frac{\pi}{2}\cos \ L, 
a_{2} = \frac{\pi}{2}$ and $b_{2} = 0$.
Thus, {\bf O1} has been achieved for this case also.

\noindent {\bf The Case $\cos  L = 0$:}
Now choose $\lambda_{1} = \frac{\pi}{2}$ and $\lambda_{2} = \frac{\pi}{2}$.
Then, the equations to solve become $\frac{a_{1}a_{2} + b_{1}b_{2}}
{\lambda_{1}\lambda_{2}} = 0$ and
$\frac {a_{2}b_{1} - a_{1}b_{2}}{\lambda_{1}\lambda_{2}}
= (-1)\sin (L)$. If $\sin (L) = 1$,
choose $a_{1} = \frac{1}{\sqrt{2}}\lambda_{1}$ and
$a_{2} = \frac{1}{\sqrt{2}}\lambda_{2}$, and $b_{1} = - a_{1}$ and
$b_{2} = a_{2}$.
If $\sin L = -1$, then choose $a_{1} = \frac{1}{\sqrt{2}}\lambda_{1}$,
$a_{2} = \frac{1}{\sqrt{2}}\lambda_{2}$, and $b_{1} = a_{1}$ and 
$b_{2} = -a_{2}$,

Now concatenating the pulses which prepare ${\mbox exp}(-iL\sigma_{y}
\otimes\sigma_{z})$ with the free evolution terms which prepare
${\mbox exp} (-i\frac{7\pi}{4}\sigma_{z}\otimes \sigma_{z})$ and
${\mbox exp} (-i\frac{\pi}{4}\sigma_{z}\otimes \sigma_{z})$
yields the stated values for the number of factors for {\bf O1}. 

To meet {\bf O2}, notice only  the
$e^{-iL \sigma_{y}\otimes \sigma_{z}}$ term needs to be addressed,
since the others
are free evolution terms and hence have pulse area equal to $0$.
Even in the preparation of the $e^{-i\L \sigma_{y}\otimes \sigma_{z}}$
term, there is only one term which is not a free evolution term (except when
$\cos L = 0$). The corresponding, $\mid b_{k}\mid$ is exactly 
equal to $ \frac{\pi}{2}\mid \sin L \mid$. So to meet {\bf O2}, 
$\theta$ has to be such that $\mid \sin L \mid \leq C\frac{2}{\pi}$.
This amounts to requiring that 
$L$ be within a prescribed bound of $0$. If $\L$ is not already
of the form, then we factor $e^{iL I_{1x}}$ as $\Pi_{k=1}^{r}
e^{iL_{k}I_{1x}}$ with the $L_{k}$ satisfying the required
deviation from $0$ condition. Clearly this can always be done.
  
Notice further, that this process of meeting {\bf O2} also ensures that
the \underline{amplitude} of the pulse, $\mid \frac{b_{k}}{a_{k}}\mid$ can also
be {\it bounded arbitrarily}. Indeed, the amplitude of the pulses are either
$0$ (corresponding to free evolution terms) or $\mid \tan L\mid$.
Clearly any process which ensures that $\mid \sin L\mid$ is within
a prescribed bound can ensure the same for $\mid \tan L\mid$.

If $\cos L = 0$,
then write ${\mbox exp}(-iLI_{1x})$
as $({\mbox exp} (-i\frac{L}{2}I_{1x}))^{2}$, and proceed as in
the $\cos L \neq 0$ cases. 
This finishes the construction.

\begin{remark}
{\rm The values obtained for $a_{k}$ and $b_{k}$ are certainly 
not the only possibilities. Since one of the principal goals of
this paper is to show that the pulse amplitudes can be arbitrarily
bounded, the $a_{k}$ and $b_{k}$ satsifying {\bf O1} were so chosen
that the resulting decomposition could be modified with minimal fuss,
to meet {\bf O2}. This means that the proof chosen was biased towards
free evolution terms. In practice, of course one can find other values
so that the cumulative time taken can also be kept within reasonable bounds.}
\end{remark}

\begin{remark}
{\rm  Suppose $S$ is prepared in the rotating frame
in $T_{S}$ units of time. Then in the orginal coordinates the matrix
$e^{-T_{S}F}S$ has been prepared. Depending on the system, it may be
desirable to rectify this error. Since, $T_{S}\geq \sum_{k=1}^{3}\theta_{k}$
[with $\theta_{k}$ given in Equation (\ref{shorteuler})],
this problem cannot be wished away by hard pulses.  Of course, hard pulses
can be further used to generate $e^{T_{S}F}$. However, this introduces
further inaccuracies in addition to those caused by the use of hard pulses
to prepare $S$ in the rotating frame.
We suggest two methods which do not need hard pulses to rectify this deviation. 
The first is to prepare $e^{T_{0}F}S$ by soft pulses
in the rotating frame for a real
parameter, $T_{0}$, to be chosen such that the time, $T_{1}$, to prepare
$e^{T_{0}F}$ in the rotating frame, satisfies $T_{1} + T_{S} = T_{0}$.
This leads to a transcendental equation for $T_{0}$
[ specifically, $\frac{1}{J}[(\frac{21}{2} +\frac{1}{ \sqrt{2}})\pi + T_{S}]
= T_{0} - \frac{\pi}{J}
(\cos \omega_{1}T_{0} + \cos\omega_{2}T_{0})$].
This has to be solved numerically.
The second method is to use optimal control \cite{herschnew} to drive
the system, in the original coordinates, from $I_{4}$ to
$e^{T_{S}F}$, with a quadratic cost
functional incorporating bounds on the
field fluence and deviation of the state from the target.
The fact that the target state, $e^{T_{S}F}$ is diagonal
will help reduce the complexity of optimal control calculations.} 
\end{remark} 
  
\end{section}

\begin{section}{Conclusions}

In this paper, a constructive procedure for generating a desired unitary
generator in $SU(4)$, via the control of a coupled, heteronuclear, two spin
system was described. The sinusoidal pulses that were produced were {\it not}
hard pulses, but instead could be bounded both in amplitude and pulse area.

An interesting problem is to investigate the methodology of this paper 
for systems which are \underline{not} studied by addressing individual
spins selectively. It is relatively straightforward to see that
an analogous Cartan decomposition can be modified to express every target
$S$ in a manner analogous to Equation (\ref{shorteuler}).
However, the remaining
calculations seem to require new methods. Investigation of this problem
will be a worthwhile research problem.

\end{section}

\begin{section} {Appendix: Determining the Euler Parameters for $SU(4)$}
To make the methodology proposed here (or for that matter any methodology
based on the Cartan decompositions) genuinely constructive, it is 
extremely desirable to find the fifteen parameters, $(D_{i}, E_{i}, F_{i}),
i = 1, \ldots , 4$ and $\theta_{k}, k = 1, \ldots , 3$ 
of Equation (\ref{ourdecomp}), as explicitly as possible,
in terms of the entries of the target matrix $S\in SU(4)$. In principle,
this amounts to solving sixteen equations for fifteen unknowns.
However, this is not really a satsifactory state of affairs. To make the
point further clear, consider determining the Euler angles of an $SU(2)$
matrix in closed form in terms of the entries of the matrix.
This is principally facilitated by the availability of the Cayley-Klein
representation, i.e., Equation (\ref{specialuseful}). A similar representation
for $SU(4)$ matrices is not available in the literature, and thus
the resulting system of equations cannot even be written down in 
a manner which will facilitate investigating the
possibility of a closed form determination of the
$D_{i}, E_{i}, F_{i}, i =1, \ldots , 4$ and the $\theta_{k}, k =1, \ldots , 3$.
 
Therefore, we eschew working directly with $S$ itself.
Instead, $S$ will be first decomposed into a product of matrices,
each of which has  
a simpler structure, so that finding
these fifteen parameters for each of these factors is more tractable.
Note that strictly speaking, this will not result in a factorization
of the form in Equation (\ref{ourdecomp}) for the original matrix $S$,
since now the factors of each of the $S_{i}$ become intertwined.
However, it will produce a factorization of the form in
Equation (\ref{shorteuler}). But, this is all that is needed to achieve
the constructive generation of $S$.

The main idea is to write $S = \Pi_{i=1}^{Q}S_{i}$ so that each of the
$S_{i}$ can be expressed solely in terms of a single $SU(2)$ matrix.
While, this may mean that the number of parameters (and, thus
the number of control pulses) needed are larger
than what would result if the fifteen parameters were directly determined
for $S$, it has the advantage that close to explicit formulae can be 
produced for the Euler angles, whereas (pending further investigation)
there is nothing
remotely close to explicit when working with $S$ itself.
Furthermore, as will be clear soon, many of the $K_{k}, k=1,\ldots ,4$
turn out to be $I_{2}$ for many of the factors $S_{i}$, and likewise
many of the $\theta_{k}$ are zero. Thus, the number of factors and pulses
is not all that high.

The decomposition that will be used is the standard Givens decomposition,
\cite{murnaghan,stewart}, modified slightly for the problem at hand.
The usual Givens decomposition is produced  
as follows. Premultiply $S^{\dagger}$, the inverse of $S$, by a sequuence
of matrices, $S_{i}$, which successively reduce the columns of $S$ to
the unit vectors, $e_{i}$, i.e., the fourth column is reduced
to $(0, 0, 0, 1)$ and the third to $(0,0,1,0)$ etc., Then it can be shown
that $Q=6$, \cite{pranew}, and thus 
\[
S = \Pi_{i=1}^{6}S_{i}
\]
Usually these $S_{i}$ are taken to be a
matrix, which upto permutation of rows and columns,
is a block matrix consisting of $I_{2}$ and an explicitly determined
$SU(2)$ matrix.
Here, we will make a slight modification, we will take $S_{6}$ to
be a tensor product of two $SU(2)$ matrices. The remaining five will be,
upto permutation, block matrices with blocks equal to $I_{2}$
and a specific $SU(2)$ matrix.

To briefly illustrate the structure of $S_{6}$ (the remaining $S_{k}$ are
constructed in the manner descirbed in \cite{pranew}), suppose that  
Suppose $S^{\dagger} = {\mbox col} (a , b ,  c ,  d )$ 
(thus $a, b, c, d$ are the four columns of $S^{\dagger}$).
 
Then choose $S_{6} = e^{i\frac{\pi}{2}\sigma_{y}}\otimes S(\alpha_{6},
\zeta_{6}, \mu_{6} )$ where $S(\alpha_{6},
\zeta_{6}, \mu_{6} )$ is the  {\it unique} $SU(2)$ matrix which takes the
vector $(d_{1}, d_{2})$ to the vector $(\mid\mid (d_{1}, d_{2})\mid\mid, 0)$.
Note that once $d_{1}, d_{2}$ are known one can explicitly write
down this $SU(2)$ matrix (here use is being made of the fact that given
any two points on a sphere of any radius in $C^{2}$ there is a unique
$SU(2)$ matrix which conveys the first to the second and that this matrix
can be found explicitly).   

The remaining, $S_{k}$ have the following structure.
\[
S_{5} = \left ( \begin{array}{cccc}
1 & 0 & 0 & 0 \\
0 & 1 &
0 & 0 \\
0 & 0 & \cos \alpha_{5}e^{i\zeta_{5}}  & \sin \alpha_{5} e^{i\mu_{5}}\\
0 & 0 &
\sin \alpha_{5} e^{i( \pi - \mu_{5})} & \cos \alpha_{5}e^{-i\zeta_{5}}  
\end{array} \right )
\]

\[
S_{4} =  \left ( \begin{array}{cccc}    
\cos \alpha_{4}e^{i\zeta_{4}} & 0 & 0 & \sin\alpha_{4} e^{i\mu_{4}} \\
0 & 1 & 0 & 0\\
0 & 0 & 1 & 0\\
\sin \alpha_{4} e^{i( \pi - \mu_{4})} & 0 & 0
&  \cos \alpha_{4}e^{-i\zeta_{4}}
\end{array} \right )
\]

\[
S_{3} = \left ( \begin{array}{cccc}
\cos \alpha_{3}e^{i\zeta_{3}} & \sin \alpha_{3} e^{i\mu_{3}} & 0 & 0 \\
\sin \alpha_{3} e^{i( \pi - \mu_{3})} & \cos \alpha_{3}e^{-i\zeta_{3}} &
0 & 0 \\
0 & 0 & 1 & 0 \\
0 & 0 & 0 & 1 \\ 
\end{array} \right )
\]

\[
S_{2} = \left ( \begin{array}{cccc}
1 & 0 & 0 & 0\\
0 & \cos \alpha_{2}e^{i\zeta_{2}} & \sin \alpha_{2} e^{i\mu_{2}} & 0 \\
0 & sin \alpha_{2} e^{i( \pi - \mu_{2})} & \cos \alpha_{2}e^{-i\zeta_{2}} & 
0 \\
0 & 0 & 0 & 1
\end{array} \right )
\]

and
\[
S_{1} = \left ( \begin{array}{cccc}
\cos \alpha_{1}e^{i\zeta_{1}} & \sin \alpha_{1} e^{i\mu_{1}} & 0 & 0 \\
\sin \alpha_{1} e^{i( \pi - \mu_{1})} & \cos \alpha_{1}e^{-i\zeta_{1}} &
0 & 0 \\
0 & 0 & 1 & 0 \\
0 & 0 & 0 & 1 \\
\end{array} \right )
\]

Once again each of the $SU(2)$ matrices, $S(\alpha_{k}, \zeta_{k}, \mu_{k} ),
k=1, \ldots , 5$, can be constructively determined from the entries of $S$. 
     
Now let us determine the fifteen Euler angles for the matrices $S_{i}, 
i = 1, \ldots , 6$. Since, the $S_{i}$ are significantly simpler than
the matrix, $S$, this is a more tractable task. In what follows the
matrices $K_{k}, i=1, \ldots, 4$ and the real constants, $\theta_{k},
k = 1, \ldots, 3$ will be described for each of the $S_{i}, i=1, \ldots, 6$.
Once the $K_{k}$ are known, it is easy to find the $\sigma_{x}, \sigma_{y}$
Euler angles, $(D_{k}, E_{k}, F_{k})$,
by using Equation (\ref{eulerangles}). Therefore, that step
will not be executed here.  

To simplify notation, three real parameters $P, Q$ and $R$  are introduced.
They are 
related to the parameters $\theta_{k}, k =1, \ldots, 3$ by:

\begin{equation}
\label{PQR}
P = \frac{\theta_{1} - \theta_{2}}{4}, \ Q = \frac{\theta_{3}}{4}, \ 
R = \frac{\theta_{1} + \theta_{2}}{4}      
\end{equation}

Below, the values of $P, Q$ and $R$ for the $S_{i}, i=1, \ldots , 6$
will be given. Obtaining the $\theta_{k}, k=1, \ldots , 3$ is then routine.
 
\noindent \underline{$S_{6}$:} Clearly, $S_{6} = K_{1}\otimes K_{2}$,
where $ K_{1} = e^{i\frac{\pi}{2}I_{1y}}$ 
and $K_{2} = e^{iD_{6}\sigma_{x}}e^{-iE_{6}\sigma_{y}}e^{-iF_{6}\sigma_{x}}$.
The Euler angles $(D_{6}, E_{6}, F_{6})$ are the Euler angles 
of the matrix  $S(\alpha_{6}, 
\zeta_{6}, \mu_{6} )$ determined according to Equation (\ref{eulerangles}).
The matrices $K_{3}, K_{4}$ can be taken to be $I_{2}$ and the constants,
$P, Q$ and $R$ (and hence  the $\theta_{k}, k=1, 2, 3$)
can be set equal to zero.

\noindent \underline{$S_{5}, S_{1}$ and $S_{3}$:}
The matrices $S_{1}$ and $S_{3}$ are essentially the same in structure,
and they are analogous to $S_{5}$. Thus, calculations for $S_{5}$
will be shown here and the modifications required for $S_{1}$ and $S_{3}$
will be given. 

So consider determing the fifteen Euler parameters of $S_{5}$. 

Pick $K_{1} = I_{2}$ and $K_{2} = S(\alpha , \zeta , \mu )$ where
the parameters $(\alpha , \zeta , \mu )$ will be presently determined.
Choose $P$ an $R$ equal to zero (thus $\alpha_{1} = 0 = \alpha_{2}$).
Multiplying out all but the factors 
$K_{3}\otimes K_{4}$, 
leads to the following matrix:
\[
\left ( \begin{array}{cccc}
\cos\alpha e^{i(\zeta - Q)} & \sin\alpha e^{i(\mu +  Q)} &
0 & 0 \\
\sin\alpha e^{i ( \pi - \mu - Q)} & \cos\alpha e^{-i (Q - \zeta_{5})} & 0
& 0 \\
0 & 0 & \cos\alpha e^{i( Q + \zeta) } & \sin\alpha e^{i(\mu - Q)}\\
0 & 0 & \sin\alpha e^{i ( \pi - \mu + Q)} 
& \cos\alpha e^{-i (Q + \zeta)}
\end{array} \right )
\]
Now choose $K_{3} = I_{2}$ and $K_{4}$ to be the inverse of the top
left hand block of the last matrix (note since the matrix in question
is in $SU(2)$ it is very straightforward to find its inverse).

This then means that
the matrix $S( \alpha_{5}, \zeta_{5}, \mu_{5} )$
(which is known) should equal the matrix:
\[
\left ( \begin{array}{cc}
\cos^{2}\alpha e^{i2Q} + sin^{2}e^{i2Q} & \cos\alpha \sin\alpha
(e^{i(2Q + \mu + \zeta - \pi)} + e^{i(\mu + \zeta - 2Q)}) \\
\cos\alpha \sin\alpha ( e^{i(\pi - \mu - \zeta + 2Q)} +
e^{-i2Q} & \sin^{2}\alpha e^{i2Q} + \cos^{2}\alpha e^{-i2Q}
\end{array} \right )
\]

We will find $\alpha, \zeta , \mu$ and $Q$ (and hence $\alpha_{3}
= 4Q$) by equating the top row of the two matrices in the
last equation (since both the matrices are in $SU(2)$, this will
automatically mean that the second row of the two are the same).
This leads to the equations:
\begin{eqnarray*}
\sqrt{\cos^{2}2Q + \cos^{2}\alpha\sin^{2}2Q} & = & \cos \alpha_{5}\\
-\cos 2\alpha\tan{2Q} & = & \tan\zeta_{5}\\
(-\frac{\cos (\mu + \zeta )}{\sin (\mu + \zeta )}) & = & \tan \mu_{5}
\end{eqnarray*}
In the above system of equations, the unknowns are $\alpha , \zeta , \mu$
and $Q$, and the known variables are $\alpha_{5}, \zeta_{5}, \mu_{5}$.
Thus, we have three transcendental equations for four unknowns. Thus, there
will be in general many solutions. Notice, that the first two equations
involve only two of the unknowns, viz., $\alpha$ and $Q$, and thus it is this
pair of equations which will have to be solved numerically. The one parameter
family of freedom comes from the third equation, where there are two unknowns.

In summary, the Euler angles of $S_{5}$ have been determined.
The Euler angles of $S_{3}$ (and, thus $S_{1}$)
can also be determined via a similar technique. Indeed, the
calculations are similar to those for $S_{3}$ except that $K_{4}$
will be taken to be the inverse of the $SU(2)$ matrix in the bottom
block of the matrix resulting from multiplying all but the last six
factors of Equation (\ref{ourdecomp}).

\noindent \underline{$S_{2}$ and $S_{4}$:}
The matrices $S_{4}$ and $S_{2}$ bear a resemblance to one another.
Calculations for $S_{2}$ will be shown together with the modifications
needed for the $S_{4}$ case. 

Choose $K_{1} = e^{i\eta_{1}\sigma_{z}}$, $K_{2} = e^{i\eta_{2}\sigma_{z}}$,
$K_{3} = I_{2}$ and $K_{4} = e^{i\eta_{3}\sigma_{z}}$, for some real
numbers $\eta_{k}, k =1, \ldots , 3$ to be determined shortly. Choose
$Q = 0$ and $P = 0$ and $R = \alpha_{2}$ (where $\alpha_{2}$ is a
Cayley-Klein parameter of $S_{2} = S (\alpha_{2}, \zeta_{2}, \mu_{2})$,
and thus, is known beforehand).

The parameters $\eta_{k}, k =1, \ldots , 3$ are found by solving the
linear system of equations:
\begin{eqnarray*}
\eta_{1} + \eta_{2} + \eta_{3} & = & 0\\
\eta_{1} - \eta_{2} + \eta_{3} & = & \zeta_{2}\\
\eta_{1} - \eta_{2} - \eta_{3} & = & \mu_{2} + \frac{\pi}{2} 
\end{eqnarray*}

For $S_{4}$ the only modification needed in this procedure is that 
$R = 0$ and $P = \alpha_{4}$.   

This completes the determination of the fifteen, $SU(4)$, ``Euler" angles for
each of the matrices $S_{i}, i=1, \ldots , 6$ in the specially chosen
Givens decomposition of the given target $S\in SU(4)$.

\begin{remark}
{\rm There is considerable liberty in the Givens decomposition. For instance,
the order in which the columns are reduced to the corresponding unit vectors
is one such degree of freedom. The factors that were chosen above were
expressly intended to facilitate the calculation of the corresponding
fifteen $SU(4)$ angles. It is an interesting problem to find other   
factorizations which yield different values of the $t_{k}$ in
Equation (\ref{shorteuler}).}
\end{remark} 
\end{section}

\end{document}